\documentclass[onecolumn,showpacs,preprintnumbers]{revtex4}
\usepackage{mathrsfs}
\usepackage{graphicx}
\usepackage{dcolumn}
\usepackage{bm}
\usepackage{epsfig}
\usepackage{amsmath,amssymb}
\usepackage{array}
\usepackage{tabularx}

\begin{document}
\title{Strong coupling constants and radiative decays of the heavy tensor mesons}
\author{Guo-Liang Yu$^{1}$}
\email{yuguoliang2011@163.com}
\author{Zhi-Gang Wang$^{1}$}
\email{zgwang@aliyun.com}
\author{Zhen-Yu Li$^{2}$}

\affiliation{$^1$ Department of Mathematics and Physics, North China
Electric power university, Baoding 071003, People's Republic of
China\\$^2$ School of Physics and Electronic Science, Guizhou Normal
College, Guiyang 550018, People's Republic of China}
\date{\today }

\begin{abstract}
In this article, we analyze tensor-vector-pseudoscalar(TVP) type of
vertices $D_{2}^{*+}D^{+}\rho$, $D_{2}^{*0}D^{0}\rho$,
$D_{2}^{*+}D^{+}\omega$, $D_{2}^{*0}D^{0}\omega$,
$B_{2}^{*+}B^{+}\rho$, $B_{2}^{*0}B^{0}\rho$,
$B_{2}^{*+}B^{+}\omega$, $B_{2}^{*0}B^{0}\omega$,
$B_{s2}^{*}B_{s}\phi$ and $D_{s2}^{*}D_{s}\phi$, in the frame work
of three point QCD sum rules. According to these analysis, we
calculate their strong form factors which are used to fit into
analytical functions of $Q^{2}$. Then, we obtain the strong coupling
constants by extrapolating these strong form factors into deep
time-like regions. As an application of this work, the coupling
constants for radiative decays of these heavy tensor mesons are also
calculated at the point of $Q^{2}=0$. With these coupling constants,
we finally calculate the radiative decay widths of these tensor
mesons.
\end{abstract}


\maketitle

\begin{large}
\textbf{1 Introduction}
\end{large}

With rapid developments of high-energy physics experiments, more and
more new states of mesons have been confirmed by D0, CDF and LHCb
collaborations\cite{Abazov1,Aaltonen1,Abazov2,Aaltonen2,Beringer,Aaij,Aaltonen3}.
The heavy-light mesons, which are composed of a heavy quark and a
light quark, can be classified into the spin doublets in the heavy
quark limit. For example, the $1S(0^{-},1^{-})$ doublets
$(B,B^{*})$, $(D,D^{*})$, $(B_{s},B_{s}^{*})$, $(D_{s},D_{s}^{*})$
and the $1P(1^{+},2^{+})$ doublets $(B_{1},B_{2}^{*})$,
$(B_{s1},B_{s2}^{*})$, $(D_{1},D_{2}^{*})$, $(D_{s1},D_{s2}^{*})$
have also been confirmed in experiments\cite{Beringer}. That is to
say, the quantum numbers $I(J^{P})$ for heavy tensor mesons
$D_{2}^{*}$, $B_{2}^{*}$, $D_{s2}^{*}$ and $B_{s2}^{*}$ are
$\frac{1}{2}(2^{+})$, $\frac{1}{2}(2^{+})$, $0(2^{+})$ and
$0(2^{+})$ respectively.

Compared with the $1S(0^{-},1^{-})$ and $1P(0^{+},1^{+})$ states of
the heavy mesons, the $1P(1^{+},2^{+})$ doublets have been drawn
little attention\cite{Swanson,Klempt}. The strong decay processes
$D_{2}^{*}\rightarrow D^{*}\pi$,
$D\pi$\cite{Abazov1,Aubert,Sanchez,Aaij2}, $D_{s2}^{*}\rightarrow
DK$\cite{Abazov1}, $B_{2}^{*}\rightarrow B^{*}\pi$,
$B\pi$\cite{Abazov1,Aaltonen2}, $B_{s2}^{*}\rightarrow BK,
B^{*}K$\cite{Aaltonen1,Abazov2,Aaij} have been observed in
experiments. In our previous work, we have studied these strong
decay processes, obtained their strong coupling constants and strong
decay widths\cite{Wang1,Li,GuoLiang}. As a continuation of these
work, we study the strong vertices $D_{2}^{*+}D^{+}\rho$,
$D_{2}^{*0}D^{0}\rho$, $D_{2}^{*+}D^{+}\omega$,
$D_{2}^{*0}D^{0}\omega$, $B_{2}^{*+}B^{+}\rho$,
$B_{2}^{*0}B^{0}\rho$, $B_{2}^{*+}B^{+}\omega$,
$B_{2}^{*0}B^{0}\omega$, $B_{s2}^{*}B_{s}\phi$ and
$D_{s2}^{*}D_{s}\phi$ and obtain its strong coupling constants.
These strong coupling constants not only play an essential role for
understanding the inner structure of these mesons but also can help
us to know about its decay behaviors. Besides, the strong coupling
constants about the heavy-light mesons can also help us
understanding the final-state interactions in the heavy quarkonium
(or meson) decays\cite{Casalbuoni,LiuX23,GuoFK}. With a fitted
function about the strong form factors in Section 3, we can also
obtain the coupling constants for the radiative decays with
intermediate momentum $Q^2=0$, which will be used to calculate the
radiative decay widths of these mesons.

To study the decay behaviors of the mesons, we can adopt several
theoretical models including perturbative and non-perturbative
methods. The QCD sum rules, proposed by Shifman, Vainshtein, and
Zakharov\cite{Shifman}, connects hadron properties and QCD
parameters\cite{Reind}. It has been widely used to study the
properties of the
hadrons\cite{Ioffe,Belyaev,Brac1,Brac2,Alie3,Alie4,Doi,Altm,Wzg3,Cerq,Rodr,Yazi,Khos1,Khos2,Rein,Pasc,Wzg5,Wang2,Navar,Khodj,GuoLY2,GuoLY4,Azizi4,Azi1,Azi2,LiuX,HeJ,LiuX2,ChenHX1,ChenHX2,Chun,ChenW,ChenW2,Zhjr1,Zhjr2,Zhjr3,Wzg}.
In this work, we analyze the tensor-vector-pseudoscalar(TVP) type of
vertices and the radiative decays using the three-point QCD sum
rules. This paper is organized as follows. After the Introduction,
we study the tensor-vector-pseudoscalar(TVP) type of strong vertices
using the three point QCD sum rules with vector mesons being
off-shell. In Sec.3, we present the numerical results and
discussions. Finally, the paper ends with the conclusions.

\begin{large}
\textbf{2 QCD sum rules for hadronic coupling constants}
\end{large}

For tensor-vector-pseudoscalar(TVP) type of vertices, its
three-point correlation function is written as,
\begin{eqnarray}
\Pi_{\mu\nu\tau}(p,p')=i^{2}\int d^{4}x\int
d^{4}ye^{i(p-p').x+ip'.y}\Big\langle0|\mathcal
{T}\Big(J_{\tau}(x)J_{\mathbb{P}}(y)J^{\dagger}_{\mu\nu}(0)\Big)
|0\Big\rangle,
\end{eqnarray}
where $J_{\mu\nu}$, $J_{\tau}$, and $J_{\mathbb{P}}$ denote
interpolating currents of heavy tensor mesons, vector mesons and
pseudoscalar mesons. These interpolating currents have the same
quantum numbers with studied mesons\cite{Ioffe,Cola},
\begin{eqnarray}
\notag
&& J_{\mu\nu}(z)=i\overline{Q}(z)\Big(\gamma_{\mu} \overleftrightarrow{D}_{\nu}+\gamma_{\nu}\overleftrightarrow{D}_{\mu}-\frac{2}{3}\widetilde{g}_{\mu\nu}\overleftrightarrow{D\!\!\!\!/}\Big)q(z) \\
\notag
&& J_{\mathbb{P}}(y)=\overline{Q}(y)i\gamma_{5}q(y)\\
\notag && J_{\tau}(x)=\overline{q}(x)i\gamma_{\tau}q(x)
\end{eqnarray}
where
\begin{eqnarray}
\notag &&
\overleftrightarrow{D}_{\mu}=(\overrightarrow{\partial}_{\mu}-ig_{s}G_{\mu})-(\overleftarrow{\partial}_{\mu}+ig_{s}G_{\mu})
\\ \notag &&
\widetilde{g}_{\mu\nu}=g_{\mu\nu}-\frac{p_{\mu}p_{\nu}}{p^{2}}
\end{eqnarray}

\begin{large}
\textbf{2.1 The hadronic side}
\end{large}

To obtain hadronic representation, we insert a complete set of
intermediate hadronic states into the correlation
$\Pi_{\mu\nu\tau}(p,p')$. These intermediate states have the same
quantum numbers with the current operators $J_{\mu\nu}$, $J_{\tau}$,
and $J_{\mathbb{P}}$. After isolating ground-state contributions of
these mesons\cite{Shifman,Reind}, the correlation function is
expressed as,
\begin{eqnarray}
\Pi^{hadr}_{\mu\nu\tau}(p,p',q)=&&\frac{\Big \langle 0|
J_{\mathbb{P}}(0)|\mathbb{P}(p')\Big \rangle \Big \langle 0|
J_{\tau}(0)|\mathbb{V}(q)\Big \rangle \Big \langle
\mathbb{T}(p)|J_{\mu\nu}^{\dag}(0)|0\Big \rangle\Big \langle
\mathbb{P}(p')\mathbb{V}(q)|L_{\mathbb{VPT}}|\mathbb{T}(p)\Big
\rangle}{(M_{\mathbb{P}}^{2}-p'^{2})(M_{\mathbb{V}}^{2}-q^{2})(M_{\mathbb{T}}^{2}-p^{2})}
+\cdots
\end{eqnarray}
The matrix elements appearing in this equation are substituted with
the following parameterized equations,
\begin{eqnarray}
\notag
&&\langle \mathbb{T}(p)| J^{\dagger}_{\mu\nu}(0)| 0\rangle=f_{\mathbb{T}}M_{\mathbb{T}}^{2}\xi^{*}_{\mu\nu}(p),\\
\notag
&&\langle0|J_{\tau}(0)|\mathbb{V}(q) \rangle=f_{\tau}M_{\tau}\zeta_{\tau}(q),\\
\notag &&\langle 0|J_{\mathbb{P}}(0)|\mathbb{P}(p')
\rangle=\frac{f_{\mathbb{P}}M_{\mathbb{P}}^{2}}{m_{Q}+m_{q}},\\
\notag && \langle
\mathbb{P}(p')\mathbb{V}(q)|L_{\mathbb{VPT}}|\mathbb{T}(p) \rangle=
g\varepsilon^{\alpha\beta\lambda\rho}p_{\alpha}\xi_{\beta\eta}p'^{\eta}q_{\lambda}\zeta_{\rho}^{*}.
\end{eqnarray}
with $q=p-p'$. Here, $f_{\mathbb{T}}$, $f_{\mathbb{P}}$ and
$f_{\tau}$ are decay constants of the tensor mesons, pseudoscalar
mesons and vector mesons, and $g$ is the strong form factor of
tensor-vector-pseudoscalar(TVP) type of vertices. Besides,
$\xi_{\mu\nu}$, $\zeta_{\rho}$ are polarization vectors of the
tensor mesons and vector mesons with the following properties,
\begin{eqnarray}
\notag &&
\xi^{*}_{\mu\nu}\xi_{\beta\eta}=\frac{1}{2}\big[\widetilde{g}_{\mu\beta}\widetilde{g}_{\nu\eta}+\widetilde{g}_{\mu\eta}\widetilde{g}_{\nu\beta}\big
]-\frac{1}{3}\widetilde{g}_{\mu\nu}\widetilde{g}_{\beta\eta}
\\ \notag &&
\zeta_{\rho}^{*}\zeta_{\tau}=g_{\rho\tau}-\frac{p_{\rho}p_{\tau}}{p^{2}}
\end{eqnarray}
With these above equations, the correlation function
$\Pi_{\mu\nu\tau}(p,p',q)$ can be expressed as follows,
\begin{eqnarray}
\notag\
&&\Pi^{hadr}_{\mu\nu\tau}(p,p',q)=\frac{gf_{\mathbb{P}}M^{2}_{\mathbb{P}}f_{\tau}M_{\tau}f_{\mathbb{T}}M_{\mathbb{T}}^{2}}{(m_{Q}+m_{q})(M_{\tau}^{2}-q^{2})(M_{\mathbb{P}}^{2}-p'^{2})(M_{\mathbb{T}}^{2}-p^{2})}\\
&&\times \Big (\frac{1}{2}p'^{\mu}\varepsilon^{\nu\tau
pp'}+\frac{1}{2}p'^{\nu}\varepsilon^{\mu\tau
pp'}-\frac{p^{2}+p'^{2}-q^{2}}{4p^{2}}p^{\mu}\varepsilon^{\nu\tau
pp'}-\frac{p^{2}+p'^{2}+q^{2}}{4p^{2}}p^{\nu}\varepsilon^{\mu\tau
pp'}\Big )+\cdots
\end{eqnarray}

\begin{large}
\textbf{2.2 The OPE side}
\end{large}

In this part, we will briefly outline the operator product
expansion(OPE) for the correlation function
$\Pi_{\mu\nu\tau}(p,p',q)$ in perturbative QCD. Firstly, we contract
all of the quark fields with Wick's theorem, and rewrite the
correlation function as follows,
\begin{eqnarray}
\Pi_{\mu\nu\tau}^{OPE}(p,p',q)=-i^{4}\int d^{4}x\int
d^{4}ye^{i(p-p').x+ip'.y}tr[S_{nm}^{q}(y-x)\gamma_{\tau}S_{mk}^{q}(x-z)\Gamma_{\mu\nu}S_{kn}^{Q}(z-y)\gamma_{5}|_{z=0}]
\end{eqnarray}
where
\begin{eqnarray}
\Gamma_{\mu\nu}=\gamma_{\mu}\overleftrightarrow{D}_{\nu}+\gamma_{\nu}\overleftrightarrow{D}_{\mu}-\frac{2}{3}\widetilde{g}_{\mu\nu}\overleftrightarrow{D\!\!\!\!/}
\end{eqnarray}
and $S^{q}$($S^{Q}$) denote light(heavy) quark propagators which can
be expressed as\cite{Pasc,Rein}.
\begin{eqnarray}
 S_{nm}^{q}(x)=&&i\frac{x\!\!\!/}{2\pi^{2}x^{4}}\delta_{nm}-\frac{m_{q}}{4\pi^{2}x^{2}}\delta_{nm}-\frac{\langle
 \overline{q}q\rangle}{12}\Big(1-i\frac{m_{q}}{4}x\!\!\!/\Big)-\frac{x^{2}}{192}m_{0}^{2}\langle
 \overline{q}q\rangle\Big( 1-i\frac{m_{q}}{6}x\!\!\!/\Big)\\
 &&
 \notag-\frac{ig_{s}\lambda_{a}^{nm}G^{a}_{\theta\eta}}{32\pi^{2}x^{2}}\Big[x\!\!\!/\sigma^{\theta\eta}+\sigma^{\theta\eta}x\!\!\!/\Big]+\cdots,
\end{eqnarray}
\begin{eqnarray}
\notag\
S_{kn}^{Q}(x)&&=\frac{i}{(2\pi )^{4}}\int d^{4}ke^{-ik.x}\bigg \{\frac{\delta _{kn}}{%
k-m_{Q}}-\frac{g_{s}G_{\alpha \beta }^{a}t_{kn}^{a}}{4}\frac{\sigma
^{\alpha\beta }(k\!\!\!/+m_{Q})+(k\!\!\!/+m_{Q})\sigma ^{\alpha
\beta }}{(k^{2}-m_{Q}^{2})^{2}}\\
& &+\frac{g_{s}D_{\alpha }G_{\beta \lambda
}^{a}t_{kn}^{a}(f^{\lambda \beta \alpha }+f^{\lambda \alpha \beta
})}{3(k^{2}-m_{Q}^{2})^{4}}
-\frac{g_{s}^{2}(t^{a}t^{b})_{kn}G_{\alpha \beta }^{a}G_{\mu \nu
}^{b}(f^{\alpha \beta \mu \nu }+f^{\alpha \mu \beta \nu }+f^{\alpha
\mu \nu \beta })}{4(k^{2}-m_{Q}^{2})^{5}}+\cdot\cdot\cdot \bigg \},
\end{eqnarray}
\begin{eqnarray}
\notag\ &&f^{\lambda \alpha \beta }=(k\!\!\!/+m_{Q})\gamma ^{\lambda
}(k\!\!\!/+m_{Q})\gamma ^{\alpha }(k\!\!\!/+m_{Q})\gamma ^{\beta
}(k\!\!\!/+m_{Q}) \\
\notag\ &&f^{\alpha \beta \mu \nu }=(k\!\!\!/+m_{Q})\gamma ^{\alpha
}(k\!\!\!/+m_{Q})\gamma ^{\beta }(k\!\!\!/+m_{Q})\gamma ^{\mu
}(k\!\!\!/+m_{Q})\gamma ^{\nu }(k\!\!\!/+m_{Q})
\end{eqnarray}
where $t^{a}=\frac{\lambda^{a}}{2}$, the $\lambda^{a}$ is the
Gell-Mann matrix, and $n$, $m$, $k$ are color indices\cite{Reind}.
In the covariant derivative, the gluon $G_{\mu}(z)$ in Eq.(4) has no
contributions as
$G_{\mu}(z)=\frac{1}{2}z^{\lambda}G_{\lambda\mu}(0)+\cdots=0$. Using
equations (4),(5), (6) and (7), the perturbative contribution of the
correlation function is written as
\begin{eqnarray}
\Pi^{pert}_{\mu\nu\tau}(p,p',q)&&=\frac{3}{(2\pi)^{4}}\int
d^{4}k\frac{tr[(k\!\!\!/+m_{q})\gamma_{\tau}(k\!\!\!/+p\!\!\!/-p'\!\!\!\!\!/+m_{q})\Gamma_{\mu\nu}(k\!\!\!/-p'\!\!\!\!\!/+m_{Q})\gamma_{5}]}{[k-m_{q}^{2}][(k+p-p')^{2}-m_{q}^{2}][(k-p')^{2}-m_{Q}^{2}]}
\end{eqnarray}
where
\begin{eqnarray}
\Gamma_{\mu\nu}=\gamma_{\mu}(2k_{\nu}-2p'_{\nu}+p_{\nu})+\gamma_{\nu}(2k_{\mu}-2p'_{\mu}+p_{\mu})
-\frac{2}{3}\Big[(g_{\mu\nu}-\frac{p_{\mu}p_{\nu}}{p^{2}})(2k\!\!\!/-2p'\!\!\!\!\!/+p\!\!\!/)
\Big ]
\end{eqnarray}
Putting all the quark lines on mass-shell by the Cutkosky¡¯s rules,
we compute the integrals both in coordinate and momentum spaces.
Then, we can obtain the spectral density by taking the imaginary
parts of the correlation function,
\begin{eqnarray}
\notag
\rho_{\mu\nu\tau}^{pert}(s,u,q^{2})=&&\frac{3}{4\pi^{2}\sqrt{\lambda}}\Big[[2A+1][A(m_{Q}-m_{q})-B(m_{q}-m_{Q})+m_{q}]\varepsilon^{\mu\tau
pp'}p_{\nu}\\
\notag
&&+[2B-2][A(m_{Q}-m_{q})-B(m_{q}-m_{Q})+m_{q}]\varepsilon^{\mu\tau
pp'}p'_{\nu} \\
\notag &&
+[2A+1][A(m_{Q}-m_{q})-B(m_{q}-m_{Q})+m_{q}]\varepsilon^{\nu\tau
pp'}p_{\mu}\\
\notag && +[2B-2][A(m_{Q}-m_{q})-B(m_{q}-m_{Q})+m_{q}]\Big
]\varepsilon^{\nu\tau pp'}p'_{\mu}+\cdots
\end{eqnarray}
where
\begin{eqnarray}
\notag
&&A=\frac{(u+m_{q}^{2}-m_{Q}^{2})(s+u-q^{2})-2u(u-q^{2}+m_{q}^{2}-m_{Q}^{2})}{\lambda(s,u,q^{2})}
\\ \notag
&&B=\frac{(u-q^{2}+m_{q}^{2}-m_{Q}^{2})(s+u-q^{2})-2s(u+m_{q}^{2}-m_{Q}^{2})}{\lambda(s,u,q^{2})}
\\ \notag
&&\lambda(s,u,q^{2})=(s+u-q^{2})^{2}-4su
\end{eqnarray}
During these derivations, we set $s=p^2$, $u=p'^2$ and $q=p-p'$ in
the spectral densities. As a result, we can see that there are
several different structures on hadronic side and OPE side. In
general, we can choose either structure to study the hadronic
coupling constant. In our calculations, we observe that the
structure $\varepsilon^{\nu\tau pp'}p_{\mu}$ can lead to pertinent
result. Using dispersion relation, the perturbative term can be
written as,
\begin{eqnarray}
\Pi_{\mu\nu\tau}^{pert}(p,p^{\prime})=\int_{s_{1}}^{s_{0}}%
\int_{u_{1}}^{u_{0}}\frac{\rho_{\mu\nu\tau}^{pert}(s,u,q^{2})}{(s-p^{2}%
)(u-p^{\prime2})}dsdu|_{-1\leq\frac{2s(m_{q}^{2}-m_{Q}^{2}+u)+(m_{Q}^{2}-m_{q}^{2}-u+q^{2})(s+u-q^{2})}{\sqrt{(m_{Q}^{2}-m_{q}^{2}-u+q^{2})^{2}-4sm_{q}^{2}}\sqrt{\lambda(s,u,q^{2})}}\leq1}
\end{eqnarray}

For non-perturbative terms, we take into account the contributions
of $\langle q\overline{q}\rangle$, $\langle
\overline{q}g\sigma.Gq\rangle$, $\langle g^{2}G^{2}\rangle$ and
$\langle f^{3}G^{3}\rangle$. After performing double Borel
transformation, we find that contributions of non-perturbative terms
come only from condensate terms $\langle g^{2}G^{2}\rangle$,
$\langle f^{3}G^{3}\rangle$. The expressions of these condensate
terms are written as,
\begin{eqnarray}
\notag  \Pi^{\langle
GG\rangle}_{\mu\nu\tau}=&&\frac{\frac{\langle\alpha_{s}GG\rangle}{\pi}}{4\pi^{2}}\Big\{[iI_{\alpha01}^{311}3m_{Q}+iI_{\alpha10}^{311}m_{Q}+iI_{\alpha01}^{411}m_{Q}^{2}(3m_{Q}-m_{q})
+iI_{\alpha10}^{411}m_{Q}^{2}(m_{Q}-m_{q})+iI_{0}^{311}m_{Q}\\
\notag &&+iI_{0}^{411}m_{Q}^{3}]
+[iI_{\alpha10}^{141}m_{q}^{2}(m_{Q}-m_{q})-iI_{\alpha01}^{131}m_{q}-iI_{\alpha01}^{141}m_{q}(m_{q}^{2}+m_{q}m_{Q})+iI_{0}^{141}m_{q}^{2}m_{Q}]
\\ \notag
&&
[iI_{\alpha10}^{114}m_{q}^{2}(m_{Q}-m_{q})+iI_{\alpha01}^{114}m_{q}^{2}(3m_{Q}-m_{q})+iI_{0}^{114}m_{q}^{2}m_{Q}-iI_{\alpha10}^{113}m_{q}
\\ \notag
&&+\frac{1}{6}[iI_{\alpha01}^{122}(m_{q}-3m_{Q})+iI_{\alpha10}^{122}(m_{q}-m_{Q})-iI_{0}^{122}m_{Q}]
\\ \notag
&&+\frac{1}{6}[iI_{\alpha01}^{221}(m_{q}-3m_{Q})-iI_{\alpha10}^{221}(m_{Q}-m_{q})-iI_{0}^{221}m_{Q}]
\\ \notag
&&+\frac{1}{6}[iI_{\alpha01}^{212}(9m_{Q}-m_{q})-iI_{\alpha10}^{212}(m_{q}-m_{Q})+iI_{0}^{212}3m_{Q}]
\end{eqnarray}
\begin{eqnarray}
\notag  \Pi^{\langle GGG\rangle}_{\mu\nu\tau}=&&\frac{\langle
g_{s}^{3}G^{a}G^{b}G^{c}f^{abc}\rangle}{12\times(2\pi)^{4}}
\Big\{[iI_{\alpha01}^{321}m_{Q}-iI_{\alpha10}^{321}(m_{q}-3m_{Q})-iI_{0}^{321}m_{Q}-iI_{\alpha10}^{421}2m_{Q}^{2}(m_{q}-m_{Q})\\
\notag
&&-iI_{\alpha01}^{421}2m_{Q}^{2}(m_{Q}+m_{q})-iI_{0}^{421}2m_{Q}^{3}]+[iI_{\alpha01}^{132}3(m_{Q}-m_{q})+iI_{\alpha10}^{132}m_{Q}+iI_{0}^{142}2m_{Q}m_{q}^{2}
\\
\notag
&&+iI_{\alpha01}^{142}2m_{q}^{2}(3m_{Q}-m_{q})+iI_{\alpha10}^{142}2m_{q}^{2}(m_{Q}-m_{q})+iI_{0}^{132}m_{Q}]+[iI_{\alpha01}^{231}(4m_{Q}-3m_{q})
\\ \notag &&
-iI_{\alpha10}^{231}(m_{q}-2m_{Q})+iI_{0}^{231}m_{Q}-iI_{0}^{141}12m_{q}-iI_{\alpha01}^{141}8m_{q}-iI_{\alpha10}^{241}2m_{q}^{2}(m_{q}-m_{Q})
\\ \notag
&&
+iI_{0}^{241}m_{Q}m_{q}(m_{q}-12m_{Q})+iI_{\alpha01}^{241}2m_{q}(4m_{q}m_{Q}-2m_{q}^{2}-4m_{Q}^{2})+[iI_{\alpha01}^{123}(3m_{Q}-2m_{q})
\\ \notag
&&+iI_{\alpha10}^{123}(m_{Q}-3m_{q})+iI_{0}^{123}m_{Q}-iN_{0}^{124}4m_{q}+iI_{\alpha01}^{124}2m_{q}(3m_{q}m_{Q}-m_{q}^{2}-4m_{Q}^{2})
\\ \notag
&&+iI_{\alpha10}^{124}2m_{q}^{2}(m_{Q}-m_{q})+iI_{0}^{124}2m_{q}m_{Q}(m_{q}-2m_{Q})+6m_{q}^{2}[iI_{\alpha10}^{115}(4m_{q}-m_{Q})\\
\notag
&&+iI_{\alpha10}^{116}6m_{q}^{2}(m_{q}-m_{Q})+iI_{\alpha01}^{115}(m_{q}+m_{Q})+iI_{\alpha01}^{116}6m_{q}^{2}(m_{q}+m_{Q})+iI_{0}^{115}m_{Q}\\
\notag &&+iI_{0}^{116}6m_{Q}m_{q}^{2}]-3[iI_{\alpha10}^{114}(4m_{q}-m_{Q})+iI_{\alpha10}^{115}6m_{q}^{2}(m_{q}-m_{Q})+iI_{\alpha01}^{114}(m_{q}+m_{Q})\\
\notag
&&+iI_{\alpha01}^{115}6m_{q}^{2}(m_{q}+m_{Q})+iI_{0}^{114}m_{Q}+iI_{0}^{115}6m_{Q}m_{q}^{2}]-\frac{3}{8}m_{q}^{2}[iI_{\alpha01}^{151}(m_{q}-4m_{Q})\\
\notag &&+iI_{\alpha01}^{161}m_{q}^{2}(m_{Q}-m_{q})-iI_{\alpha10}^{151}(m_{q}-3m_{Q})-iI_{\alpha10}^{161}6m_{q}^{2}(m_{q}-3m_{Q})+iI_{0}^{151}m_{Q}\\
\notag &&+iI_{0}^{161}6m_{q}^{2}m_{Q}]+\frac{3}{16}[iI_{\alpha01}^{141}(m_{q}-4m_{Q})+iI_{\alpha01}^{151}m_{q}^{2}(m_{Q}-m_{q})-iI_{\alpha10}^{141}(m_{q}-3m_{Q})\\
\notag
&&-iI_{\alpha10}^{151}6m_{q}^{2}(m_{q}-3m_{Q})+iI_{0}^{141}m_{Q}+iI_{0}^{151}6m_{q}^{2}m_{Q}]+\frac{3}{16}[iI_{\alpha01}^{511}6m_{Q}^{2}(3m_{Q}-m_{q})\\
\notag
&&-iI_{\alpha01}^{411}(m_{q}-12m_{Q})-iI_{\alpha10}^{511}6m_{Q}^{2}(m_{q}-m_{Q})-iI_{\alpha10}^{411}(m_{q}-4m_{Q})+iI_{0}^{411}4m_{Q}\\
\notag &&+iI_{0}^{511}6m_{Q}^{3}]-\frac{3}{8}m_{Q}^{2}[iI_{\alpha01}^{611}6m_{Q}^{2}(3m_{Q}-m_{q})-iI_{\alpha01}^{511}(m_{q}-12m_{Q})+iI_{0}^{511}4m_{Q}\\
\notag
&&-iI_{\alpha10}^{611}6m_{Q}^{2}(m_{q}-m_{Q})-iI_{\alpha10}^{511}(m_{q}-4m_{Q})+iI_{0}^{611}6m_{Q}^{3}]+[iI_{\alpha01}^{222}(3m_{Q}-m_{q})\\
\notag &&
-iI_{\alpha10}^{222}(m_{q}-3m_{Q})+iI_{0}^{222}m_{Q}]\Big\}
\end{eqnarray}
where
\begin{eqnarray}
\notag\
 I_{0}^{abc}&=&\frac{(-1)^{a+b+c}\pi
^{2}i}{\Gamma (a)\Gamma (b)\Gamma
(c)(M_{1}^{2})^{b}(M_{2}^{2})^{c}(M^{2})^{a-2}}\int_{0}^{\infty
}d\tau (\tau +1)^{a+b+c-4}\tau ^{1-b-c}\\
&&\exp \{-\frac{1}{\tau
}\frac{Q^{2}}{M_{1}^{2}+M_{2}^{2}}-\frac{(\tau
+1)m_{Q}^{2}}{M^{2}}-\frac{(\tau +1)m_{q}^{2}}{\tau
M_{1}^{2}}-\frac{(\tau +1)m_{q}^{2}}{\tau M_{2}^{2}}\}
\end{eqnarray}
\begin{eqnarray}
\notag\
 I_{\alpha01}^{abc}&=&\frac{(-1)^{a+b+c}\pi
^{2}i}{\Gamma (a)\Gamma (b)\Gamma
(c)(M_{1}^{2})^{b}(M_{2}^{2})^{c+1}(M^{2})^{a-3}}\int_{0}^{\infty
}d\tau (\tau +1)^{a+b+c-5}\tau ^{1-b-c}\\
&&\exp \{-\frac{1}{\tau
}\frac{Q^{2}}{M_{1}^{2}+M_{2}^{2}}-\frac{(\tau
+1)m_{Q}^{2}}{M^{2}}-\frac{(\tau +1)m_{q}^{2}}{\tau
M_{1}^{2}}-\frac{(\tau +1)m_{q}^{2}}{\tau M_{2}^{2}}\}
\end{eqnarray}
\begin{eqnarray}
\notag\
 I_{\alpha10}^{abc}&=&\frac{(-1)^{a+b+c}\pi
^{2}i}{\Gamma (a)\Gamma (b)\Gamma
(c)(M_{1}^{2})^{b+1}(M_{2}^{2})^{c}(M^{2})^{a-3}}\int_{0}^{\infty
}d\tau (\tau +1)^{a+b+c-5}\tau ^{1-b-c}\\
&&\exp \{-\frac{1}{\tau
}\frac{Q^{2}}{M_{1}^{2}+M_{2}^{2}}-\frac{(\tau
+1)m_{Q}^{2}}{M^{2}}-\frac{(\tau +1)m_{q}^{2}}{\tau
M_{1}^{2}}-\frac{(\tau +1)m_{q}^{2}}{\tau M_{2}^{2}}\}
\end{eqnarray}

\begin{large}
\textbf{3 The results and discussions}
\end{large}

Estimating the parameters of the lowest-lying hadronic state are in
general plagued by the presence of unknown subtraction terms, the
spectral function of excited and continuum states. This situation
can be substantially improved by applying to both OPE side and
phenomenological side the Borel transformation\cite{Cola}. Thus, we
perform the double Borel transform with respect to the variables
$P^2=-p^2$, $P'^2=-p'^2$ and match OPE side with the hadronic
representation Eq.(3), invoking the quark-hadron duality. Finally,
we obtain the QCDSR as follows,
\begin{eqnarray}
 && \notag
\frac{gf_{\mathbb{P}}M^{2}_{\mathbb{P}}f_{\tau}M_{\tau}f_{\mathbb{T}}M_{\mathbb{T}}^{2}}{(m_{q}+m_{Q})(M_{\tau}^{2}+Q^{2})}
\Big(-\frac{M_{\mathbb{T}}^{2}+M_{\mathbb{P}}^{2}-Q^{2}}{4M_{\mathbb{T}}^{2}}\Big)\frac{1}{M_{1}^{2}M_{2}^{2}}exp[-\frac{M_{\mathbb{T}}^{2}}{M_{1}^{2}}]exp[-\frac{M_{\mathbb{P}}^{2}}{M_{2}^{2}}]
\\ \notag  = && \frac{1}{M_{1}^{2}M_{2}^{2}}\int_{s_{1}}^{s_{0}} \int_{u_{1}}^{u_{0}}
exp[-\frac{s}{M_{1}^{2}}]exp[-\frac{u}{M_{2}^{2}}]\rho^{pert}(s,u,Q^{2})dsdu
\\ &&
 +\Pi^{\langle
GG\rangle}_{\mu\nu\tau}(M_{1}^{2},M_{2}^{2},Q^{2})+\Pi^{\langle
GGG\rangle}_{\mu\nu\tau}(M_{1}^{2},M_{2}^{2},Q^{2})
\end{eqnarray}
Here, $Q^2=-q^2$, parameters $s_{0}$ and $u_{0}$ are used to further
reduce the contributions from excited and continuum states. Its
values are employed as $s_{0}=(m_{i}+\Delta_{i})^2$ and
$u_{0}=(m_{o}+\Delta_{o})^2$, where $m_{i}$ and $m_{o}$ are ground
state masses of the in-coming and out-coming hadron. In general,
$\Delta_{i}$ and $\Delta_{o}$ are expected to be $0.3GeV\sim0.5GeV$,
which can guarantee the values of $s_{0}$ and $u_{0}$ be close to
the mass squared of the first excited state of these in-coming and
out-coming hadrons\cite{Wang1}. Parameters $M_1^2$ and $M_2^2$ in
Eq.(14) are Borel parameters. In order to determine optimal values
about these above parameters, two criteria should be considered.
First, pole contribution should be as large as possible comparing
with contributions of higher and continuum states. Secondly, we
should also ensure OPE convergence and the stability of our results.
That is to say, the results which are extracted from sum rules,
should be independent of the Borel parameters. One can consult
Ref.\cite{Li,GuoLiang} for technical details of these processes. As
for the other parameters in Eq.(14), their values are all listed in
Table 1.
\begin{table*}[t]
\begin{ruledtabular}\caption{Input parameters used in this analysis.}
\begin{tabular}{c c c c c c}
  Parameters & \ Values($MeV$ )    &\ Parameters &\ Values &\ Parameters &\ Values \\
\hline
$m_{D_{2}^{+*}}$    &  \   $2465.4\pm1.3$  \cite{Tanabashi}  &\ $m_{B_{s2}^{*}}$      &  \   $5839.85 \pm 0.17$ \cite{Tanabashi} &\ $\langle \frac{\alpha_{s}G^{2}}{\pi}\rangle$        &  \    $(0.012\pm0.004)$  $GeV^4$  \cite{Narison1,Narison2,Narison3}           \\
$m_{D_{2}^{0*}}$     &  \    $2460.7\pm0.4$ \cite{Tanabashi}   &\ $m_{B_{s}}$      &  \   $5366.89 \pm 0.19$  \cite{Tanabashi} &\  $\langle g_{s}^{3}GGG\rangle$        &  \  $(0.045\pm0.002)GeV^6$  \cite{Narison1,Narison2,Narison3}      \\
$m_{D^{+}}$       &  \      $1869.65 \pm 0.05$ \cite{Tanabashi}    &\ $m_{\rho}$      &  \   $775.26 \pm 0.25$ \cite{Tanabashi} &\ $f_{\phi}$        &  \    $(229\pm3)  MeV$  \cite{Tanabashi}        \\
$m_{D^{0}}$      &  \   $1864.83 \pm 0.05$  \cite{Tanabashi}  &\  $m_{\omega}$      &  \   $782.65 \pm 0.12$ \cite{Tanabashi} &\ $f_{\rho}$        &  \     $(210\pm4) MeV$ \cite{Tanabashi}   \\
$m_{B_{2}^{+*}}$      &  \   $5737.2 \pm 0.7$  \cite{Tanabashi}    &\  $m_{\phi}$      &  \   $1019.46 \pm 0.016$ \cite{Tanabashi} &\ $f_{\omega}$        &  \    $197\pm8 MeV $ \cite{Bharucha}   \\
$m_{B_{2}^{0*}}$     &  \   $5739.5 \pm 0.7$ \cite{Tanabashi} &\  $m_{u}$      &  \   $2.2^{+0.5}_{-0.4}$ \cite{Tanabashi} &\ $f_{D}$        &  \   $203.7\pm4.7$ $MeV$   \cite{Tanabashi}     \\
$m_{B^{+}}$      &  \   $5279.32 \pm 0.14$ \cite{Tanabashi} &\  $m_{d}$      &  \   $4.7^{+0.5}_{-0.3}$  \cite{Tanabashi} &\ $f_{B_{2}^{*}}$        &  \    $110\pm11$  $MeV$  \cite{Wang22}   \\
$m_{B^{0}}$      &  \   $5279.63 \pm 0.15$ \cite{Tanabashi}   &\  $m_{b}$      &  \   $4180^{+25}_{-35}$ \cite{Tanabashi} &\ $f_{D_{2}^{*}}$        &  \    ($182\pm20$)    $MeV$  \cite{Wang22}     \\
$m_{D_{s2}^{*}}$      &  \   $2569.1 \pm 0.8$  \cite{Tanabashi}   &\  $m_{c}$      &  \   $1275\pm 25$ \cite{Tanabashi} &\ $f_{B}$   & \ $188\pm25MeV$\cite{Tanabashi}               \\
$m_{D_{s}}$      &  \   $1968.3 \pm 0.07$ \cite{Tanabashi} &\ $m_{s}$ & \ $95^{+9}_{-3}$ \cite{Tanabashi}  &\  $f_{B_{s2}^{*}}$         &  \  $134\pm11MeV$\cite{Wang22} \\
$f_{D_{s2}^{*}}$  & \  $222\pm21$ \cite{Wang22} & \ $f_{B_{s}}$ & \
$231\pm16$ \cite{Wang22} & \ $f_{D_{s}}$ & \ $257.8\pm4.1MeV$ \cite{Tanabashi}\\
\end{tabular}
\end{ruledtabular}
\end{table*}

The strong form factor $g$ from Eq.(14) are obtained in deep
space-like region $q^2 \rightarrow-\infty$, where the intermediate
mesons are off-shell. In order to obtain strong coupling constants,
we must extrapolate these results into deep time-like region. It is
noticed that there are no exact expressions for the dependence of
the strong form factors on $Q^2$. In this work, we find that the
results can be fitted into the exponential function,
\begin{equation}
g(Q^{2})=Aexp[{BQ^2}]
\end{equation}

\begin{figure}[h]
\begin{minipage}[h]{0.45\linewidth}
\centering
\includegraphics[height=5cm,width=7cm]{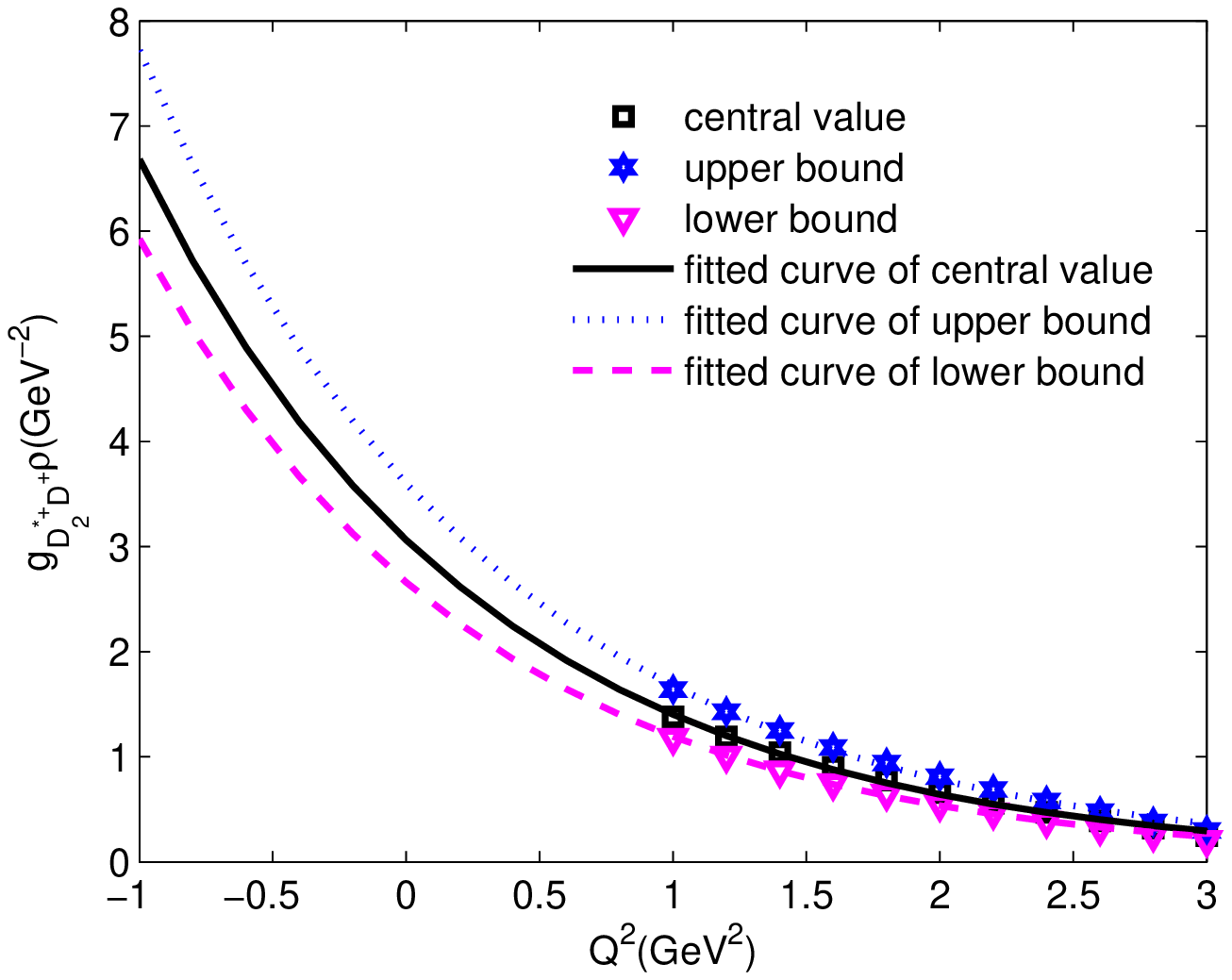}
\caption{The strong form factor $g_{D_{2}^{*+}D^{+}\rho}$, and its
fitted results as a function of $Q^2$.\label{your label}}
\end{minipage}
\hfill
\begin{minipage}[h]{0.45\linewidth}
\centering
\includegraphics[height=5cm,width=7cm]{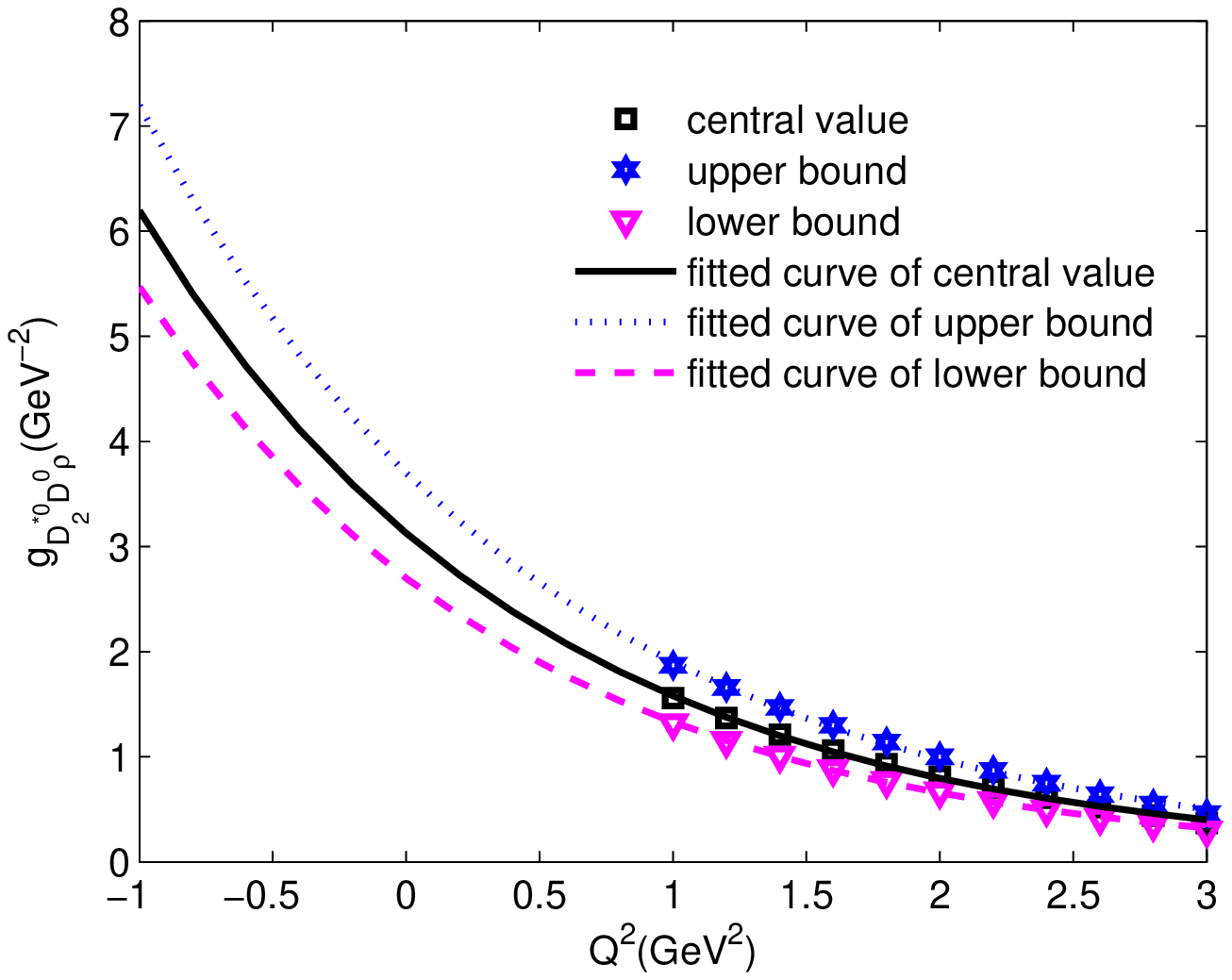}
\caption{The strong form factor $g_{D_{2}^{*0}D^{0}\rho}$, and its
fitted results as a function of $Q^2$.\label{your label}}
\end{minipage}
\end{figure}
\begin{figure}[h]
\begin{minipage}[h]{0.45\linewidth}
\centering
\includegraphics[height=5cm,width=7cm]{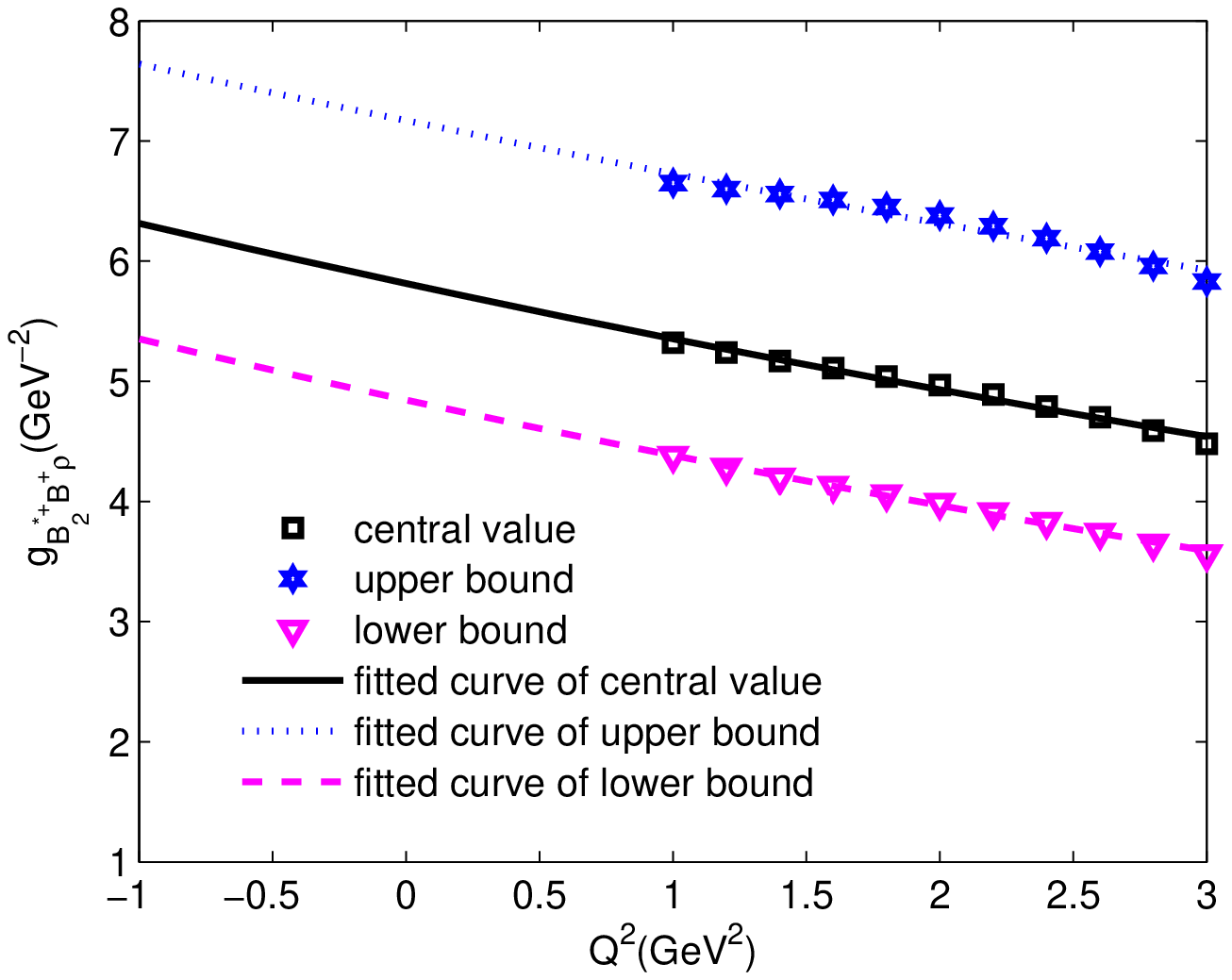}
\caption{The strong form factor $g_{B_{2}^{*+}B^{+}\rho}$, and its
fitted results as a function of $Q^2$.\label{your label}}
\end{minipage}
\hfill
\begin{minipage}[h]{0.45\linewidth}
\centering
\includegraphics[height=5cm,width=7cm]{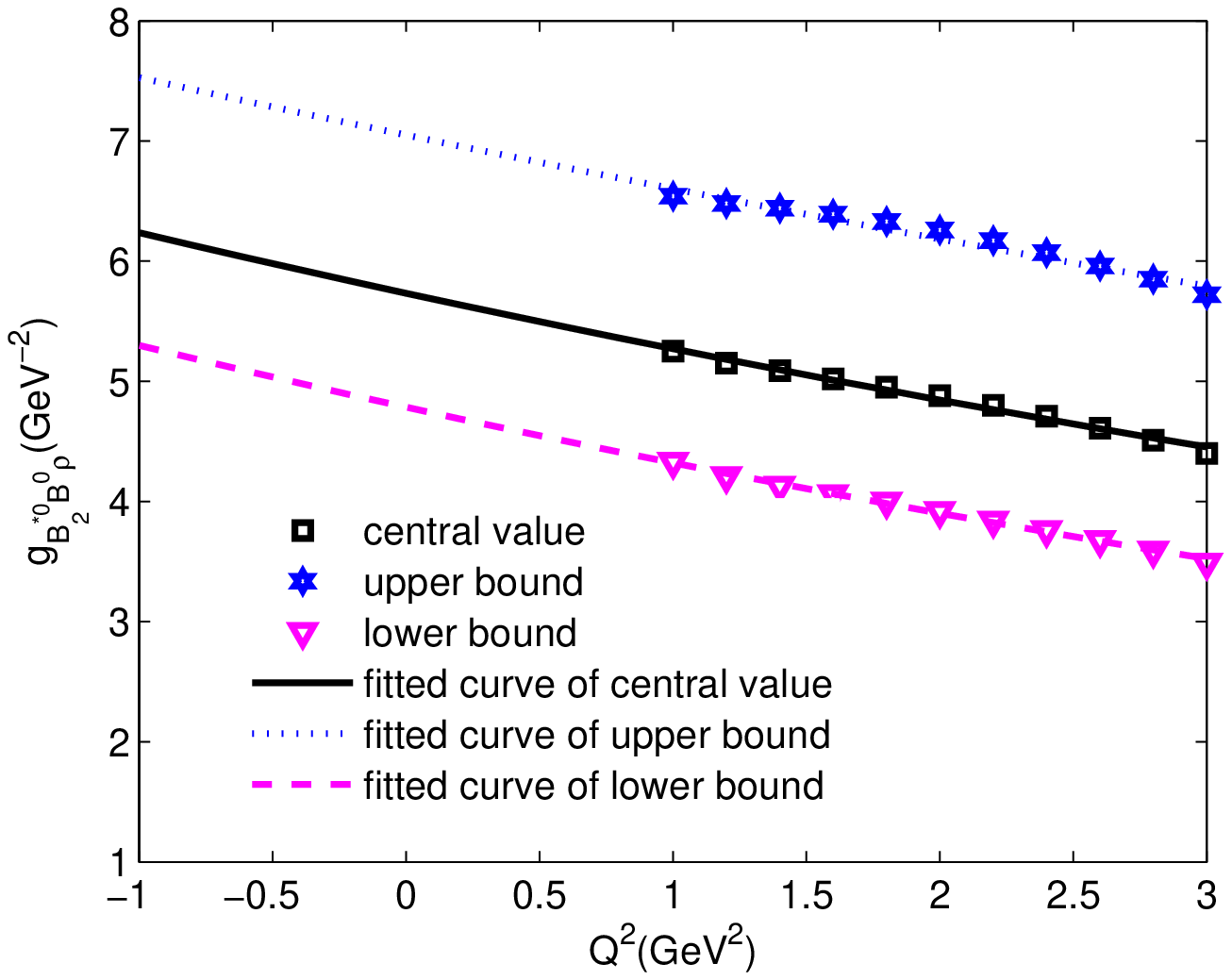}
\caption{The strong form factor $g_{B_{2}^{*0}B^{0}\rho}$, and its
fitted results as a function of $Q^2$.\label{your label}}
\end{minipage}
\end{figure}
\begin{figure}[h]
\begin{minipage}[h]{0.45\linewidth}
\centering
\includegraphics[height=5cm,width=7cm]{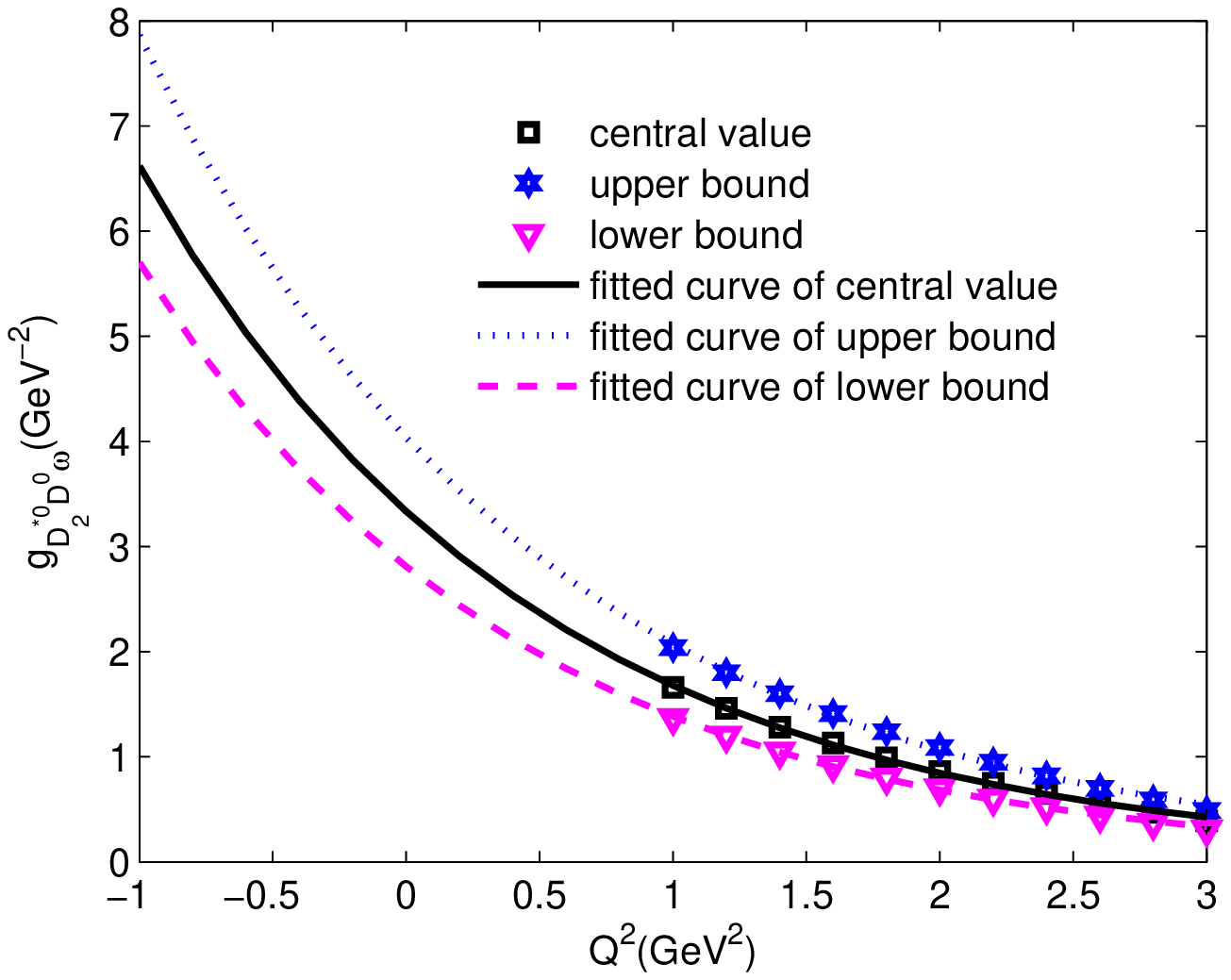}
\caption{The strong form factor $g_{D_{2}^{*0}D^{0}\omega}$, and its
fitted results as a function of $Q^2$.\label{your label}}
\end{minipage}
\hfill
\begin{minipage}[h]{0.45\linewidth}
\centering
\includegraphics[height=5cm,width=7cm]{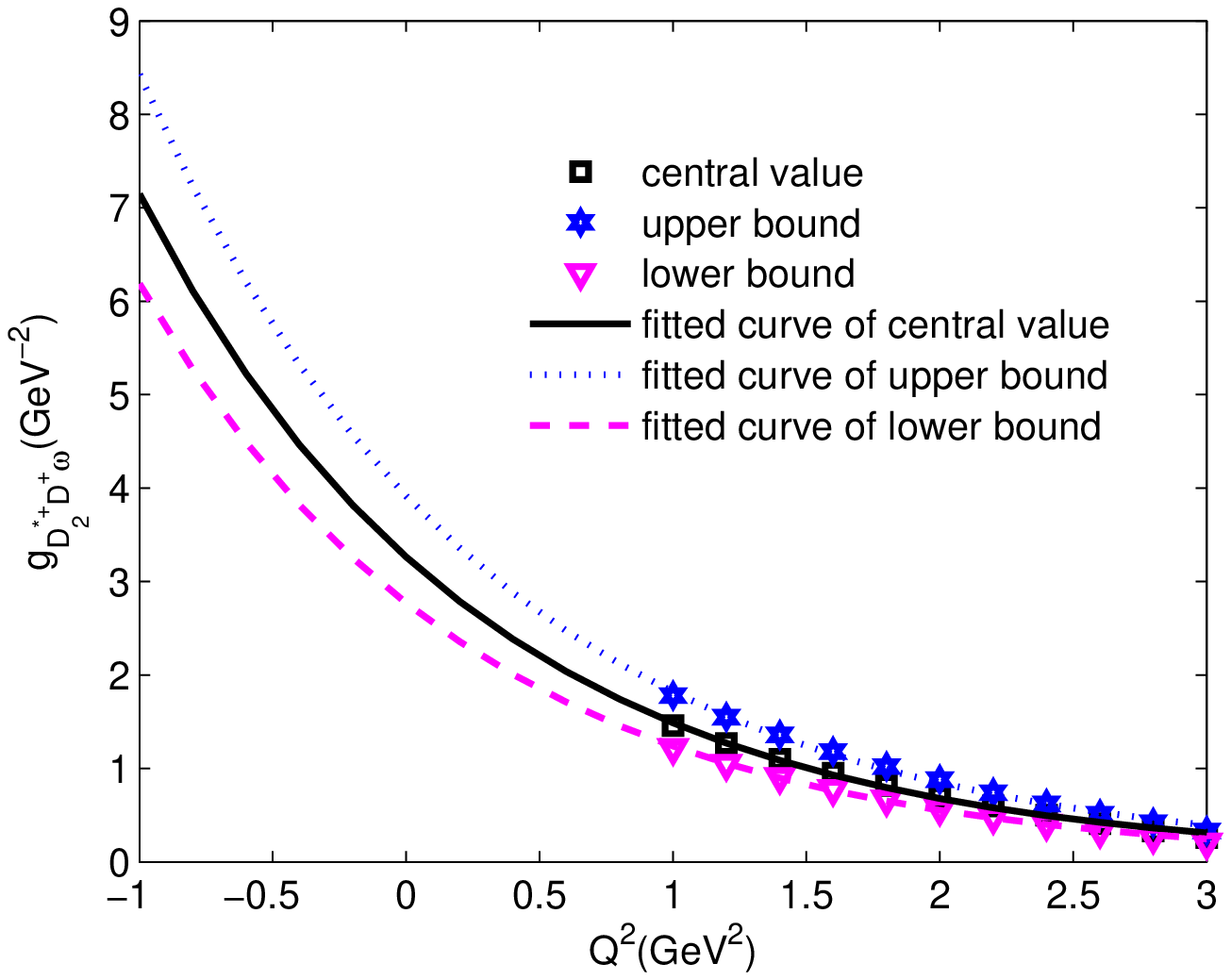}
\caption{The strong form factor $g_{D_{2}^{*+}D^{+}\omega}$, and its
fitted results as a function of $Q^2$.\label{your label}}
\end{minipage}
\end{figure}
\begin{figure}[h]
\begin{minipage}[h]{0.45\linewidth}
\centering
\includegraphics[height=5cm,width=7cm]{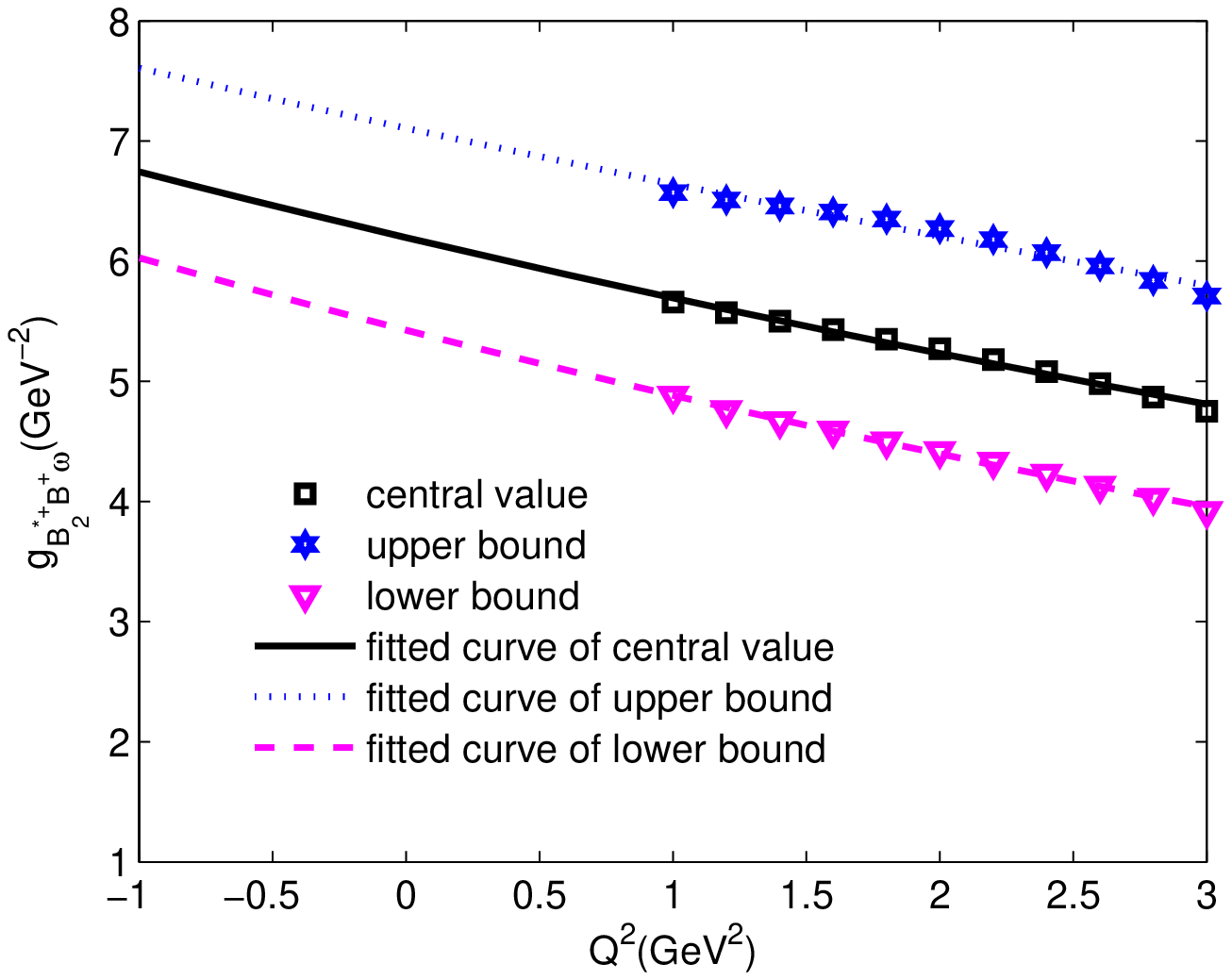}
\caption{The strong form factor $g_{B_{2}^{*+}B^{+}\omega}$, and its
fitted results as a function of $Q^2$.\label{your label}}
\end{minipage}
\hfill
\begin{minipage}[h]{0.45\linewidth}
\centering
\includegraphics[height=5cm,width=7cm]{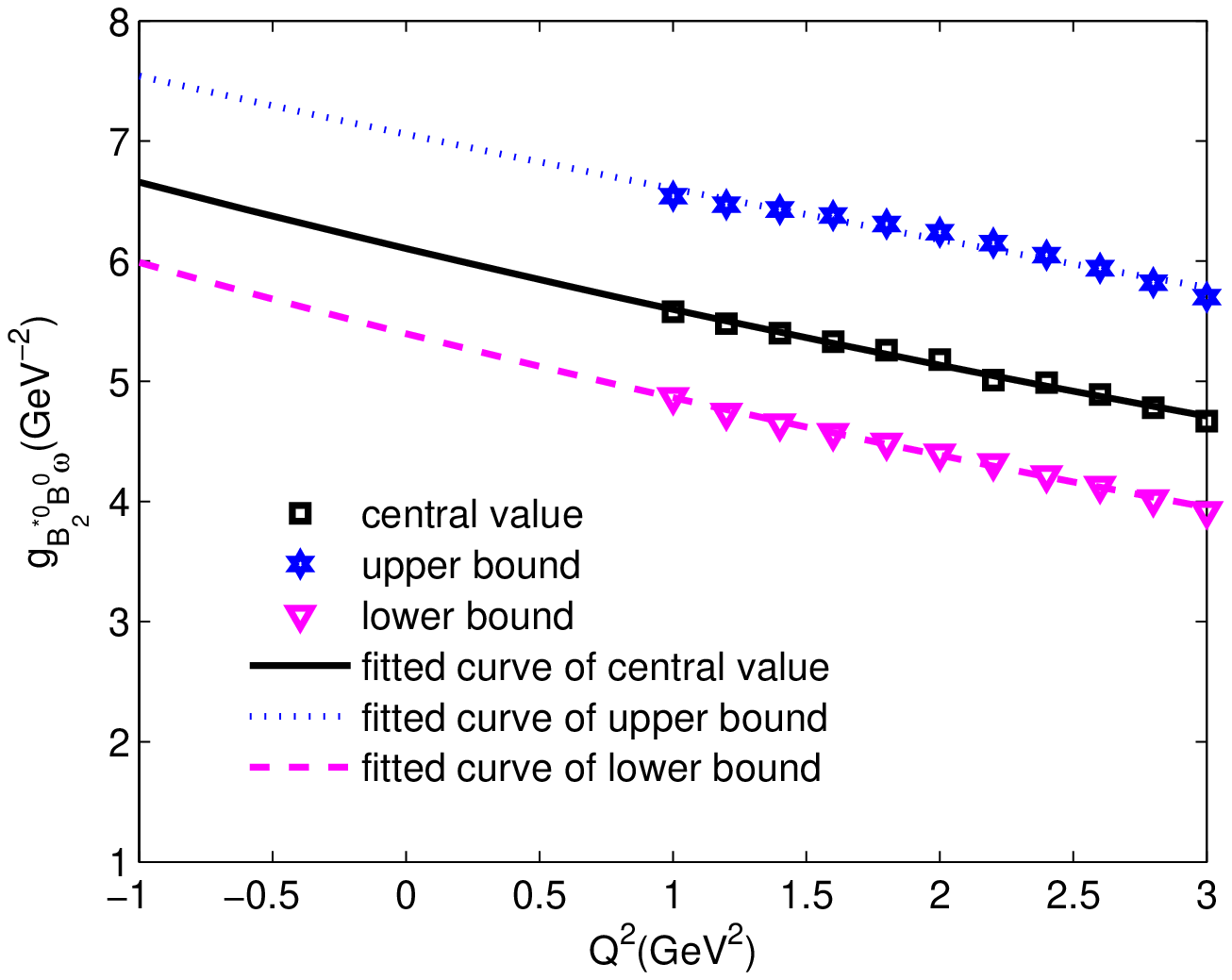}
\caption{The strong form factor $g_{B_{2}^{*0}B^{0}\omega}$, and its
fitted results as a function of $Q^2$.\label{your label}}
\end{minipage}
\end{figure}
\begin{figure}[h]
\begin{minipage}[h]{0.45\linewidth}
\centering
\includegraphics[height=5cm,width=7cm]{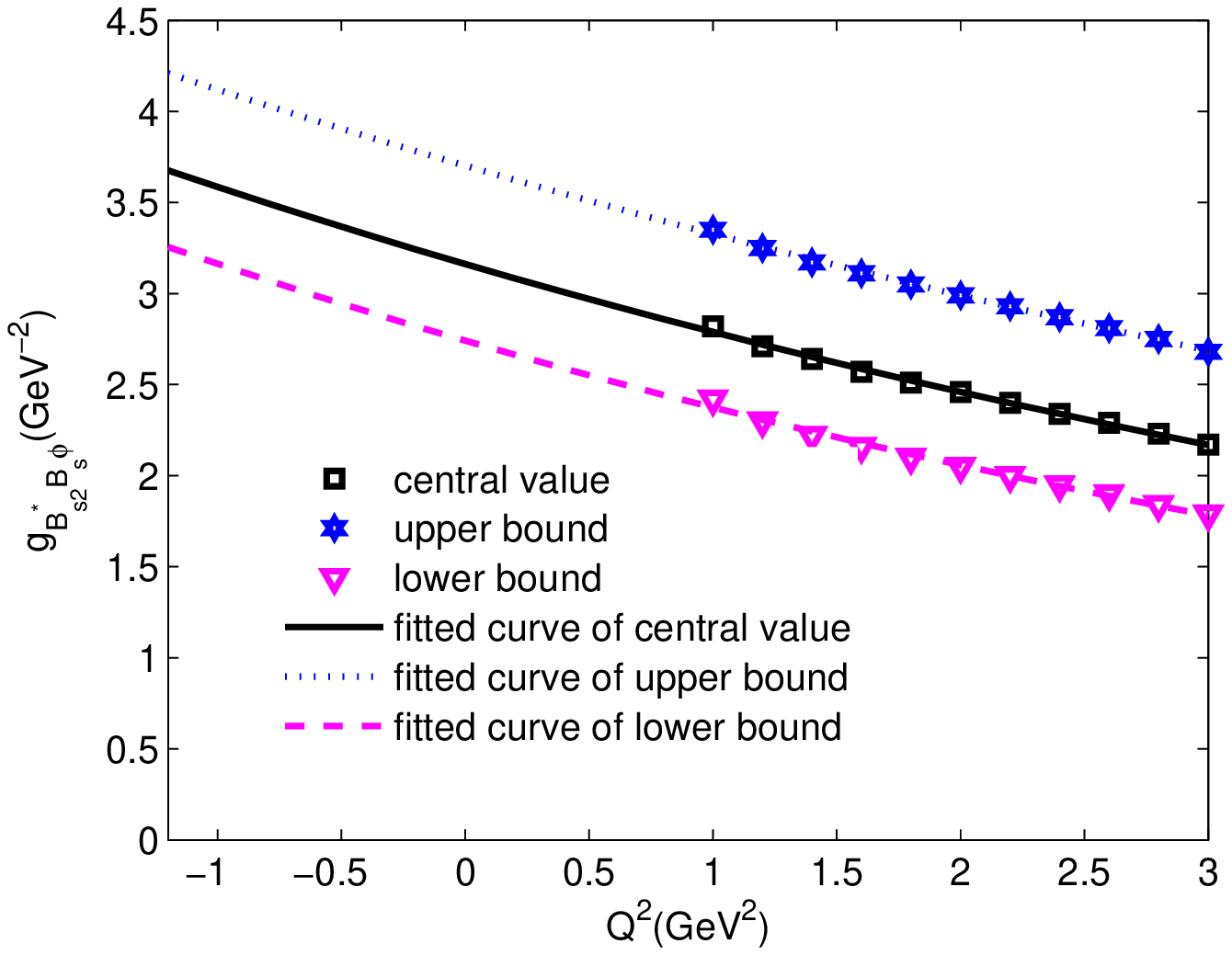}
\caption{The strong form factor $g_{B_{s2}^{*}B_{s}\phi}$, and its
fitted results as a function of $Q^2$.\label{your label}}
\end{minipage}
\hfill
\begin{minipage}[h]{0.45\linewidth}
\centering
\includegraphics[height=5cm,width=7cm]{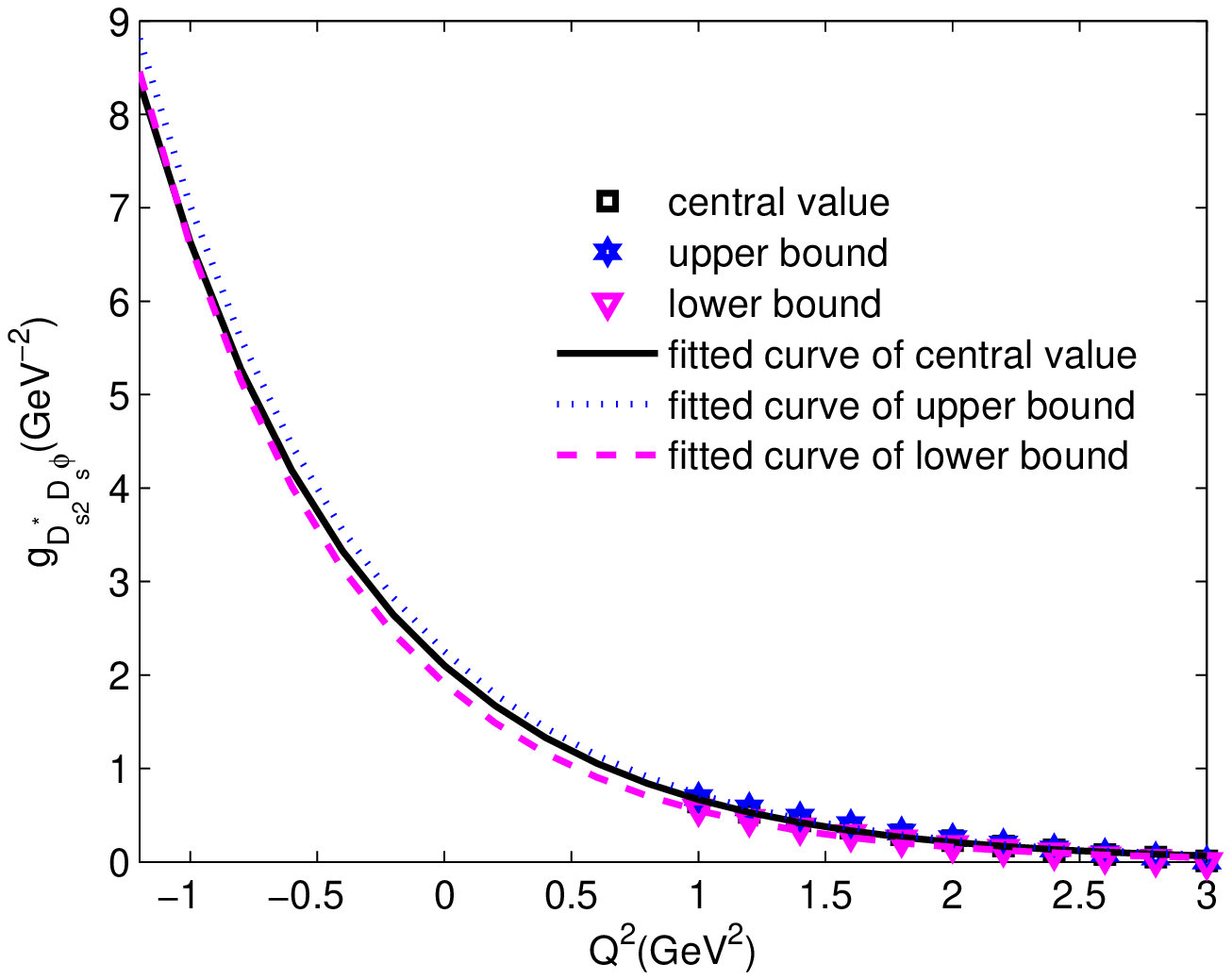}
\caption{The strong form factor $g_{D_{s2}^{*}D_{s}\phi}$, and its
fitted results as a function of $Q^2$.\label{your label}}
\end{minipage}
\end{figure}

\begin{table*}[t]
\begin{ruledtabular}\caption{Strong coupling constants and fitted parameters about the central value, upper bound and lower bound.}
\begin{tabular}{c c c c c c c c}
 Mode     &\ g($GeV^{-2}$) & \ $A_{c}$   & \ $B_{c}$ &\ $A_{u}$  & \ $B_{u}$ &\ $A_{l}$ &\ $B_{l}$ \\
\hline
$D_{2}^{*+}$$\rightarrow$ $D^{+}\rho$    &  \  $4.89^{+0.80}_{-0.41}$  & \ 3.062 & \  -0.781   &\  3.596     &\  -0.764  &\ 2.659 &\ -0.801 \\
 $D_{2}^{*0}$$\rightarrow$ $D^{0}\rho$    &  \    $4.71^{+0.80}_{-0.41}$   &  \   3.127 &\ -0.684 &\ 3.698 &\ -0.665 &\ 2.698 &\ -0.706   \\
$B_{2}^{*+}$$\rightarrow$ $B^{+}\rho$        &  \   $6.11^{+1.34}_{-0.97}$   &  \   5.814 &\ -0.083 &\ 7.169 &\ -0.064 &\ 4.844 &\ -0.100   \\
$B_{2}^{*0}$$\rightarrow$ $B^{0}\rho$       &  \   $6.03^{+1.30}_{-0.94}$   &  \   5.733 &\ -0.084 &\ 7.049 &\ -0.065 &\ 4.785 &\ -0.102   \\
$D_{2}^{*0}$$\rightarrow$ $D^{0}\omega$    &  \  $5.07^{+1.00}_{-0.74}$  & \ 3.333 & \  -0.686   &\  4.034     &\  -0.667  &\ 2.810 &\ -0.707 \\
 $D_{2}^{*+}$$\rightarrow$ $D^{+}\omega$    &  \    $5.28^{+0.98}_{-0.75}$   &  \   3.262 &\ -0.785 &\ 3.915 &\ -0.757 &\ 2.769 &\ -0.804   \\
$B_{2}^{*+}$$\rightarrow$ $B^{+}\omega$        &  \   $6.53^{+0.88}_{-0.74}$   &  \   6.196 &\ -0.084 &\ 7.107 &\ -0.068 &\ 5.426 &\ -0.105   \\
$B_{2}^{*0}$$\rightarrow$ $B^{0}\omega$       &  \   $6.44^{+0.91}_{-0.69}$   &  \   6.105 &\ -0.086 &\ 7.056 &\ -0.066 &\ 5.397 &\ -0.104   \\
$B_{s2}^{*}$$\rightarrow$ $B_{s}\phi$        &  \   $3.60^{+0.54}_{-0.42}$   &  \   3.162 &\ -0.126 &\ 3.702 &\ -0.107 &\ 2.741 &\ -0.143   \\
$D_{s2}^{*}$$\rightarrow$ $D_{s}\phi$       &  \   $6.93^{+0.40}_{-0.02}$   &  \   2.102 &\ -1.149 &\ 2.254 &\ -1.136 &\ 1.907 &\ -1.241   \\
\end{tabular}
\end{ruledtabular}
\end{table*}

In Figs.1-10, we show the values of the strong form factors on $Q^2$
which are obtained from Eq.(14) and the fitting curve, in which it
is marked as Central value and Fitted curve of Central value. Thus,
we can obtain the strong coupling constants by taking
$Q^{2}=-m_{on-shell}^2$ for intermediate mesons in the fitting
function Eq.(15). The values of fitted parameters $A$ and $B$ in
Eq.(15) and the strong coupling constants are all listed in Table
II.

The uncertainties of strong form factors in Eq.(14) mainly come from
input parameters
$m_{D_{2}^{*+}}$,$m_{D_{2}^{0*}}$,$m_{u}$,$m_{b}$,$f_{\rho}$,
$f_{\phi}$, $\langle \overline{q}q\rangle$, $\cdots$ Theoretically,
we can calculate its values with uncertainty transfer formula
$\delta=\sqrt{\Sigma_{i}(\frac{\partial f}{\partial
x_{i}})^{2}(x_{i}-\overline{x}_{i})^{2}}$, where $f$ denotes the
strong form factor in Eq.(14), and $x_{i}$ denotes input parameters.
For simplicity, the upper and lower limits of the results are
estimated by taking $f^{upper(lower)}=f(\overline{x}_{i}\pm\Delta
x_{i})$, which are marked as Upper bound and Lower bound in
Figs.1-10. After these approximations, they are also fitted into
exponential functions and are also extrapolated into the physical
regions in order to get the uncertainties of the strong coupling
constants. These results are all listed in Table II.

Finally, we give an analysis of the radiative decays of the heavy
tensor mesons $\mathbb{T}\rightarrow \mathbb{P}\gamma$. The coupling
constants of these radiative decays $g_{\mathbb{T}\mathbb{P}\gamma}$
can be easily obtained by setting $Q^2=0$ in Eq.(15). The radiative
decay width can be expressed as the following representation,
\begin{eqnarray}
  \Gamma=\frac{1}{2J+1}\sum \frac{|p|}{8\pi M_{i}^{2}}|T|^{2}
\end{eqnarray}
\begin{eqnarray}
\notag
p=\frac{\sqrt{[M_{i}^{2}-(M_{f}+m)^{2}][M_{i}^{2}-(M_{f}-m)^{2}]}}{2M_{i}}
\end{eqnarray}
where $i$ and $f$ denote the initial and final state mesons, $J$ is
the total angular momentum of the initial meson, $\sum$ denotes the
summation of all the polarization vectors, and $T$ denotes the
scattering amplitudes. The radiative decays $\mathbb{T}\rightarrow
\mathbb{P}\gamma$ can be described by the following electromagnetic
lagrangian $\pounds$
\begin{eqnarray}
\notag\ \pounds=-eQ_{q}\overline{q}\gamma_{\mu}qA^{\mu}
\end{eqnarray}
From this lagrangian, the decay amplitude can be written as,
\begin{eqnarray}
\notag\  T&=&\left\langle \mathbb{P}(p')\gamma (q,\varepsilon
)|\mathbb{T}(p,\xi_{\beta\eta}^{\omega})\right\rangle \\
\notag\
&=&\left\langle \gamma (q,\varepsilon )|\mathbb{V}(q,\zeta_{\rho})\right\rangle \frac{i}{%
q^{2}-m_{\mathbb{V}}^{2}}\left\langle \mathbb{P}(p')\mathbb{V}
(q,\zeta_{\rho} )|\mathbb{T}(p,\xi_{\beta\eta}^{\omega} )\right\rangle \\
\notag\ &=&\left\langle \mathbb{P}(p')\mathbb{V} (q,\zeta_{\rho}
)|\mathbb{T}(p,\xi_{\beta\eta}^{\omega} )\right\rangle
\frac{i}{q^{2}-m_{\mathbb{V} }^{2}}f_{\mathbb{V} }m_{\mathbb{V}
}eQ_{q}(-i)\varepsilon _{\kappa }^{*}\zeta
^{\kappa }\\
&=&g_{\mathbb{T}\mathbb{P}\gamma}\epsilon ^{\alpha \beta \lambda
\rho }p_{\alpha }\xi _{\beta\eta }p'^{\eta}q_{\lambda }\zeta _{\rho
}^{\ast }\frac{i}{q^{2}}f_{\mathbb{V} }m_{\mathbb{V}
}eQ_{q}(-i)\varepsilon _{\kappa }^{*}\zeta ^{\kappa }
\end{eqnarray}
Here, $p_{\alpha}$, $p'^{\eta}$ and $q_{\lambda}$ are the four
momenta of the tensor meson, pseudoscalar meson and $\gamma$, $\xi$,
$\zeta$ and $\varepsilon$ are their polarization vectors,
respectively. With Eqs.(16) and (17),  we can obtain the radiative
decay width of $\mathbb{T}\rightarrow \mathbb{P}\gamma$,
\begin{eqnarray}
  \Gamma=\frac{1}{10}\alpha
Q_{[u,d,s]}^{2}g_{\mathbb{T}\mathbb{P}\gamma}^{2}|\frac{f_{\mathbb{V}}}{m_{\mathbb{V}}}|^{2}[\frac{M_{T}^{2}-M_{P}^{2}}{2M_{T}}]^{3}\{\frac{1}{6}[\frac{5M_{T}^{2}-2M_{P}^{2}}{2M_{T}}]^{2}-\frac{2}{3}M_{T}^{2}\}
\end{eqnarray}
where
$\alpha=\frac{1}{137}$,$Q_{u}=\frac{2}{3}$,$Q_{d}=Q_{s}=-\frac{1}{3}$.
Considering different decay channels, we obtain the widths of
different radiative decays which are listed in Table III. From
reference\cite{Tanabashi}, we can see the decay widths of the tensor
mesons, $\Gamma(D_{2}^{*0})=47.5\pm1.1MeV$,
$\Gamma(D_{2}^{*\pm})=46.7\pm1.2MeV$,
$\Gamma(D_{s2}^{*})=16.9\pm0.8MeV$,
$\Gamma(B_{2}^{*0})=24.2\pm1.7MeV$, $\Gamma(B_{2}^{*+})=20\pm5MeV$,
$\Gamma(B_{s2}^{*})=1.47\pm0.33MeV$. From these experimental data,
we observe that the branching ratios of the calculated radiative
decays are of the order of $10^{-2}\sim10^{-5}$, which are
measurable in the future by LHCb. In reference\cite{Aliev10}, the
radiative decays of the heavy tensor mesons were also analyzed in
the framework of the light cone QCD sum rules method. We observe
that our results for mesons $D_{2}^{*}$ and $D_{s2}^{*}$ are
comparable with its results. For mesons $B_{2}^{*}$ and
$B_{s2}^{*}$, the results from QCD sum rules and light cone QCD sum
rules vary widely, which need to be further studied by other
theoretical methods or in experiments.

\begin{table*}[tb]
\caption{The decay widths for different radiative decays.}
\begin{tabularx}{8.3cm}{p{4cm}<{\centering} p{4cm}<{\centering}}
\hline \hline
Radiative decay     &\ $\Gamma$($keV$)  \\
\hline
$D_{2}^{*+}$$\rightarrow$ $D^{+}\gamma$    &  \  $0.464^{+0.126}_{-0.079}$ \\
 $D_{2}^{*0}$$\rightarrow$ $D^{0}\gamma$    &  \  $2.730^{+0.776}_{-0.491}$   \\
 $D_{s2}^{*}$$\rightarrow$ $D_{s}\gamma$       &  \   $1.658^{+0.198}_{-0.010}$   \\
$B_{2}^{*+}$$\rightarrow$ $B^{+}\gamma$        &  \   $1108^{+315}_{-202}$   \\
$B_{2}^{*0}$$\rightarrow$ $B^{0}\gamma$       &  \   $275^{+78.1}_{-48.0}$   \\
$B_{s2}^{*}$$\rightarrow$ $B_{s}\gamma$        &  \   $37.6^{+12.2}_{-8.2}$   \\
\hline \hline
\end{tabularx}
\end{table*}

\begin{large}
\textbf{4 Conclusion}
\end{large}

In this paper, we analyze the tensor-vector-pseudoscalar(TVP) type
of vertices in the cases of light vector mesons $\rho$, $\omega$ and
$\phi$ being off-shell. We firstly calculate its strong form factors
in space-like regions($q^{2}<0$). Then, we fit the form factors into
exponential functions which are used to extrapolate into time-like
regions($q^{2}>0$) to obtain strong coupling constants. These strong
coupling constants are important parameters in studying the strong
decay behaviors of the tensor mesons in the future. Setting
intermediate momentum $Q^2=0$ in the fitted analytical functions
about strong form factors, we also obtained the coupling constants
of the radiative decays of the tensor mesons. With these coupling
constants, we calculate the radiative decay widths of these tensor
mesons and compare our results with experimental data and those of
other research groups.


\begin{large}
\textbf{Acknowledgment}
\end{large}

This work has been supported by the Fundamental Research Funds for
the Central Universities, Grant Number $2016MS133$, Natural Science
Foundation of HeBei Province, Grant Number $A2018502124$.

\end{document}